\documentclass[%
reprint,
amsmath,amssymb,
aip,
jap,
]{revtex4-2}

\usepackage{graphicx}
\usepackage{dcolumn}
\usepackage{bm}
\usepackage[mathlines]{lineno}

\begin{document}

	\title{Modulation of Magnetic Anisotropy and Spin-Orbit Interaction by Electrical Current in FeCoB Nanomagnets }
	
	\author{Vadym Zayets}
	\affiliation{National Institute of Advanced Industrial Science and Technology (AIST), Umezono 1-1-1, Tsukuba, Ibaraki, Japan}
	
	\email{v.zayets@aist.go.jp; v.zayets@gmail.com}
	\homepage{https://www.zayets2physics.com/}

	\date{\today}


\begin{abstract}
We present a novel method for measuring the modulation of magnetic anisotropy and the strength of spin-orbit interaction by an electrical current in nanomagnets. Our systematic study explores the current dependencies of these properties across a variety of nanomagnets with different structures, compositions, and sizes, providing unprecedented insights into the complex physical origins of this effect. We identified two distinct contributions to the observed current modulation: one proportional to the current and the other to the square of the current. The squared-current contribution, originating from the Spin Hall effect, uniquely accumulates strength with an increasing number of interfaces, resulting in exceptionally large current modulation of magnetic anisotropy and spin-orbit interaction in multi-layer nanomagnets. Conversely, the linear-current contribution stems from the Ordinary and Anomalous Hall effects and exhibits opposite polarity at different interfaces, making it significant only in asymmetrical single-layer nanomagnets. The squared-current contribution induces substantial anisotropy field changes, up to 30-50$\%$ at typical MRAM recording currents, leading to thermally-activated magnetization reversal and data recording. The linear-current contribution, while smaller, is effective for parametric magnetization reversal, providing sufficient modulation for efficient data recording through resonance mechanisms. This finding highlights the complex nature of spin accumulation and spin dynamics at the nanoscale, presenting an opportunity for further optimization of data recording in MRAM technology.

\end{abstract}



\maketitle


\section{Introduction. Two types of MRAM recording}

The primary current challenge in the development of Magnetic Random Access Memory (MRAM) technology is reducing the energy required for data recording. In MRAM technology, data is recorded in a memory cell through the application of an electrical pulse, which reverses the magnetization direction of a nanomagnet \cite{MRAM2020Everspin,MRAM2017Ohno,MRAMCubukcu}. This reversal allows the nanomagnet to switch between two stable magnetization states, effectively storing a bit of data \cite{ZayetsArch2019Hc}. The reduction of recording current is crucial for the efficient and reliable functioning of MRAM.

There are two main types of MRAM, each with a different recording mechanism: Spin-Transfer Torque MRAM (STT-MRAM) and Spin-Orbit Torque MRAM (SOT-MRAM).

In STT-MRAM \cite{MRAM2020Everspin,MRAM2017Ohno,MRAMCubukcu,STT_Slonczewski}, the magnetization reversal occurs when spin-polarized conduction electrons are driven by an electrical current from another ferromagnetic electrode to the nanomagnet. For reliable magnetization reversal, the number of injected spin-polarized electrons must exceed a certain threshold, making it challenging to reduce the recording current for this mechanism.

Conversely, in SOT-MRAM \cite{SOT_Miron2011,SOT_Ralph2011,SOT_Sinova2020}, an electrical current creates a spin accumulation and spin depletion \cite{SpinIject_Crooker2005SpinInjectionImage,Zayets2020MishenkoSpinPol} at the opposite boundaries of the nanomagnet, leading to magnetization reversal and data recording. Unlike STT-MRAM, the spin-polarized electrons in SOT-MRAM are generated at the boundary within the same nanomagnet rather than being injected from an external source. This intrinsic generation allows for optimization of the recording mechanism, enabling a significant reduction in the recording current compared to STT-MRAM. Additionally, highly efficient resonance-type magnetization reversal mechanisms, such as parametric reversal induced by the magnetic field created by spin accumulation \cite{ZayetsJMMM2023Parametric}, can be employed in SOT-MRAM.

When spins with differing orientations are injected into a nanomagnet, the resulting change in spin distribution causes the overall spin of the nanomagnet to rotate. This process is the primary mechanism behind STT and SOT types of magnetization reversal. Additionally, there exists another mechanism by which current-induced spin accumulation can influence and even reverse magnetization. The spin-polarized electrons influence the strength of the spin-orbit interaction and, consequently, the magnetic anisotropy of the nanomagnet. Magnetic anisotropy is crucial for maintaining two stable magnetization directions, which are essential for data storage. When the recording electrical pulse reduces magnetic anisotropy, it facilitates the magnetization reversal process, thereby enhancing the efficiency and reliability of data recording, or even triggering the magnetization reversal by itself.

This paper elaborates on how spin accumulation, generated by an electrical current, affects magnetic anisotropy and the strength of the spin-orbit interaction. It also explores how this effect can be utilized for effective MRAM recording.

\section{Measurement of spin- orbit interaction strength}

Recently, a novel method for measuring the strength of spin-orbit interaction has been introduced \cite{Zayets2024SObasic}, offering valuable insights into this complex fundamental phenomenon. This measurement technique provides profound experimental insights of the physical processes that influence spin-orbit interaction and, consequently, govern magnetic anisotropy. The modulation of spin-orbit interaction strength by gate voltage \cite{ZayetsArch2024_SO_VCMA}  and the dependence of spin-orbit interaction on interface polarity \cite{ZayetsArch2024_SO_interface} were experimentally observed. The current paper presents a systematic investigation into the modulation of spin-orbit strength by an electrical current, alongside an exploration of the mechanisms through which the electrical current affects strength of the spin-orbit interaction. The purpose of this study is to elucidate how the modulation of magnetic anisotropy and spin-orbit interaction can facilitate the magnetization reversal process for MRAM recording.

The measurement technique is based on a measurement of the strength of magnetic anisotropy under an external magnetic field applied along the magnetic easy axis. The external magnetic field amplifies the spin-orbit (SO) interaction, thereby enhancing magnetic anisotropy. Since magnetic anisotropy arises from SO interaction, the increased anisotropy measured under an increased external magnetic field allows for the evaluation of the strength of SO interaction.

\begin{figure}[tb]
	\begin{center}
		\includegraphics[width=7 cm]{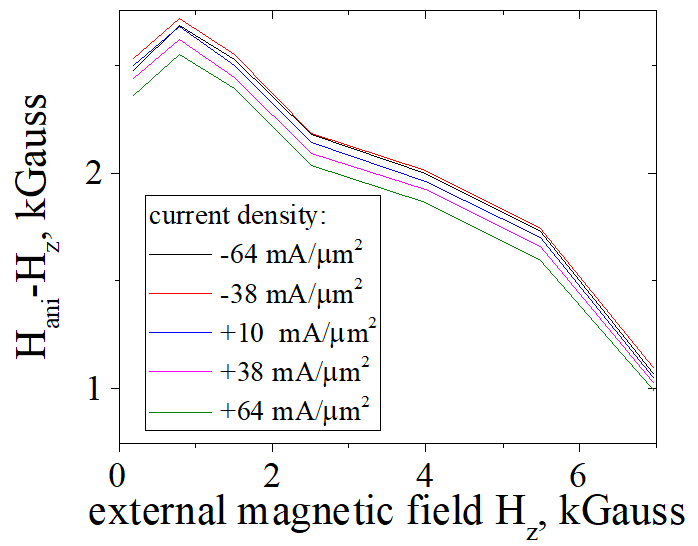}
	\end{center}
	\caption{
		{Anisotropy field $H_{ani} $ as a function of external perpendicular-to-plane magnetic field $H_{z} $ measured at a different current density. Sample $Ta(2.5)/ FeB(1.1)/MgO(6)$}. Number in blankets describes layer thickness in nanometers. 
	}
	\label{fig:FigHaniHz} 
\end{figure}


The parameter that quantifies magnetic anisotropy is the anisotropy field $H_{ani}$.  Both theoretical evaluations and experimental measurements have demonstrated that $H_{ani}$  increases linearly with the external magnetic field $H_z$  applied along the magnetic easy axis. This relationship can be expressed \cite{Zayets2024SObasic} as:

\begin{equation}
	H_{ani}-H_z= H^0_{ani}+k_{so}H_z
	\label{EqHaniHz}
\end{equation} 
where $k_{so}$ is the coefficient of spin- orbit interaction, which defines the strength of spin-orbit interaction, and $H^0_{ani}$ is the anisotropy field in absence of $H_z$. 

\section{Experimental details}

To measure the current-dependency of the strength of spin-orbit interaction, a nanomagnet was fabricated atop a Ta or W nanowire, with an attached Hall probe aligned to the nanomagnet. Two types of nanomagnets were investigated: a single-layer nanomagnet composed of FeB or FeCoB, and a multi-layer nanomagnet consisting of alternating FeB and non-magnetic Ta layers. Each nanomagnet was coated with a MgO layer. Nanomagnets of varying sizes, ranging from 50 nm x 50 nm to 2000 nm x 2000 nm, were fabricated at different locations on a single wafer.

Experiments were conducted at room temperature, well below the Curie temperature of FeB and FeCoB. The measurement procedure involved measuring the Hall angle while sweeping an in-plane external magnetic field $H_x$. A perpendicular-to-plane magnetic field $H_z$ was used as a parameter and  maintained constant during the $H_x$ sweep. The magnetic field $H_x$ tilts the magnetization of the nanomagnet. The tilt angle was determined by measuring the reduction in Hall voltage.  The anisotropy field $H_{ani}$ was evaluated by fitting the linear relationship between the measured in-plane magnetization component  $M_x$ and and  $H_x$. Using the evaluated value of $H_{ani}$,  the coefficient of spin-orbit interaction $k_{so}$ was determined from the slope of the linear dependency between  $H_{ani}$ and  $H_z$. All details of the measurement procedure are described in Ref. \cite{Zayets2024SObasic}.

\begin{figure}[tb]
	\begin{center}
		\includegraphics[width=8.5 cm]{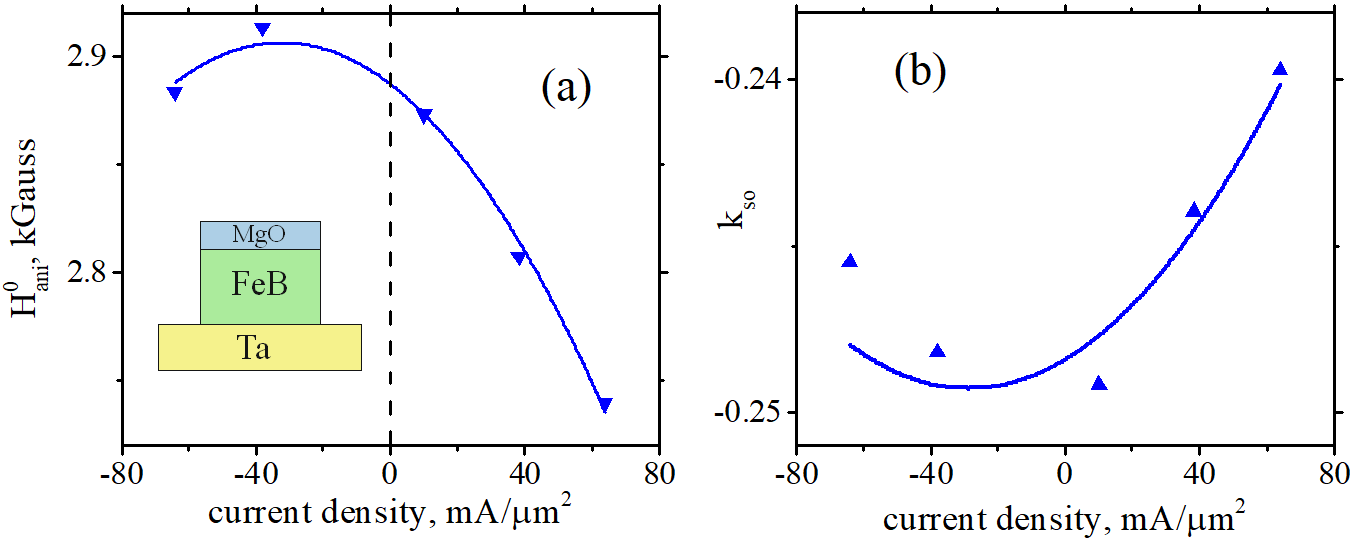}
	\end{center}
	\caption{
		{Anisotropy field $H^0_{ani}$ and coefficient of the spin orbit interaction $k_{so}$ as a function of the current density measured in single-layer nanomagnet: $Ta(2.5)/ FeB(1.1)/MgO(6)$} 
	}
	\label{fig:FigHaniKsoCurrentSingle} 
\end{figure}


\section{Disparity in Current Dependency of Spin- Orbit interaction : Single-Layer versus Multi-Layer Nanomagnets }

Figure \ref{fig:FigHaniHz} shows the measured relationship between $H_{ani}-H_z$ and the external magnetic field $H_z$  in a single-layer nanomagnet measured for different current densities. The data shows an approximately linear trend with minor oscillations superimposed, which are a known characteristic of the spin-orbit interaction in the interfacial layer \cite{Zayets2024SObasic}. Two key features are evident from the measurements. The first feature is that the offset of each line consistently decreases as the current density increases. The second feature is that the gap between the lines narrows at higher $H_z$, indicating that the slope of the lines increases with increasing current density. As seen in Eq. (\ref{EqHaniHz}), the slope determines $k_{so}$  and the offset determines $H^0_{ani}$. These parameters exhibit opposite dependencies on the current density.

Figure \ref{fig:FigHaniKsoCurrentSingle} illustrates the measured dependence of  $k_{so}$ and $H^0_{ani}$  on the current density j for the same single-layer nanomagnet. In both cases, there are two nearly equal contributions: one linearly proportional to the current density (~j) and the other proportional to the square of the current density (~$j^2$). For both $k_{so}$ and $H^0_{ani}$, the linear and square contributions have opposite polarities.

Figure \ref{fig:FigHaniKsoCurrentMulti} shows a similar current dependence of $k_{so}$ and $H^0_{ani}$, but for a multilayer nanomagnet. There are striking differences between the current dependencies for single-layer and multilayer nanomagnets. Firstly, the current-induced changes in both  $k_{so}$ and $H^0_{ani}$ are an order of magnitude larger for the multilayer nanomagnet compared to the single-layer nanomagnet. At the same current density, the changes in $H^0_{ani}$ and $k_{so}$ are about 0.1 kGauss and 0.01 for the single-layer nanomagnet, respectively, but approximately 1 kGauss and 0.4 for the multilayer nanomagnet. Secondly, the square- current contribution is substantially larger than the linear- current contribution for the multilayer nanomagnet, in contrast to their nearly equal contributions in the single-layer nanomagnet. This trend is systematic and was observed consistently across all studied nanomagnets.

\section{General Trends in Distribution Across Different Nanomagnets}

\begin{figure}[tb]
	\begin{center}
		\includegraphics[width=7.2 cm]{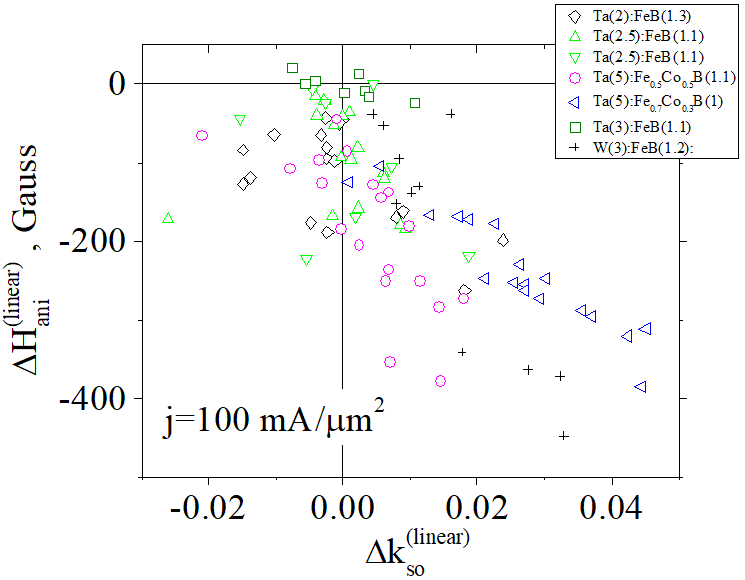}
	\end{center}
	\caption{
		{The change of anisotropy field  $\Delta H_{ani}^{(linear)}$  versus the change of spin-orbit interaction coefficient $\Delta k_{so}^{(linear)}$ as current density changes from 0 to 100 $mA/\mu m^2$ , measured in single-layer nanomagnets. Only the components linearly proportional to the current are shown. Each dot represents an individual nanomagnet measurement. Dots of the same color and shape correspond to nanomagnets fabricated at different locations on the same wafer. The number in brackets indicates the layer thickness in nanometers.} 
	}
	\label{fig:FigDistributionLinear} 
\end{figure}

In Figure \ref{fig:FigDistributionLinear}, the current- induced change of $H^0_{ani}$ are plotted against the current- induced changed of $k_{so}$ as current density is switched to 100 $mA/\mu m^2$, measured in single-layer nanomagnets of various sizes and structures. Only the components linearly proportional to the current are shown. There is some variation in the data even for nanomagnets fabricated on the same wafer, due to slight differences in surface roughness and thickness across different areas of the wafer.

\begin{figure}[tb]
	\begin{center}
		\includegraphics[width=8.5 cm]{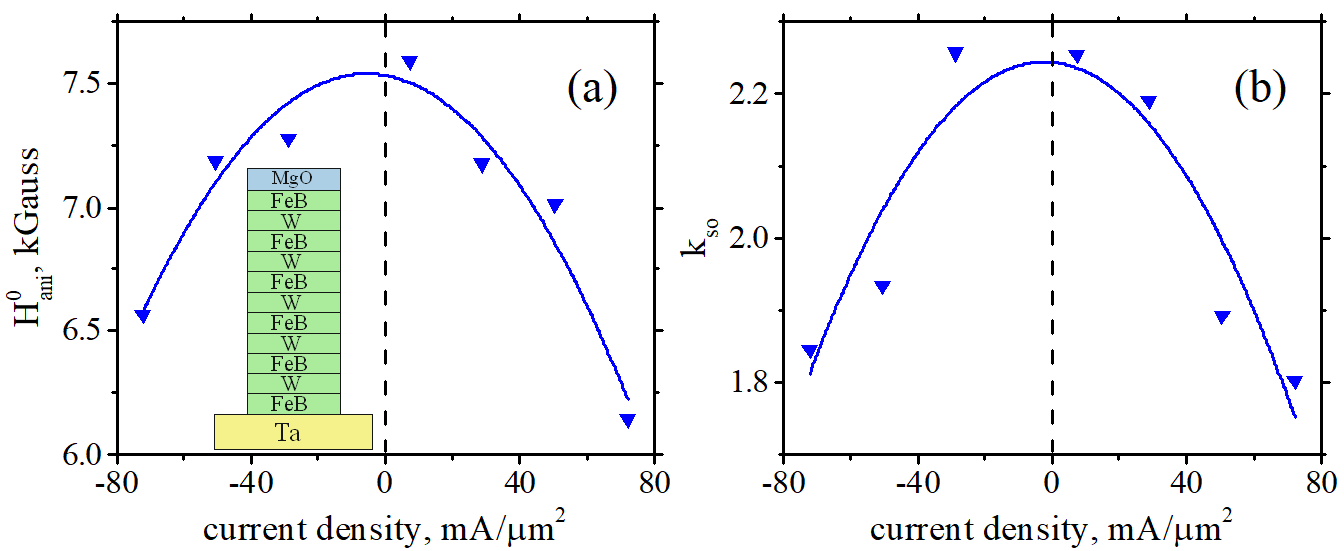}
	\end{center}
	\caption{
		{Anisotropy field $H^0_{ani}$ and coefficient of the spin orbit interaction $k_{so}$ as a function of the current density measured in multi-layer nanomagnet: $W(3)/ [FeB(0.55)/ W(0.5)]_5/FeB(0.55)/MgO(5.5)$} 
	}
	\label{fig:FigHaniKsoCurrentMulti} 
\end{figure}


The typical recording current density in present MRAM ranges between 20 and 60 $mA/\mu m^2$.  Consequently, the current-induced change of $H_{ani}$ shown in Figure \ref{fig:FigDistributionLinear} represents the largest expected variation of anisotropy field  at realistic MRAM recording currents, which is relatively small. The measured anisotropy field for the same nanomagnets ranged from 2 to 6 kGauss, indicating that the current-induced change in the anisotropy field does not exceed 10$\%$ and is insufficient to cause magnetization reversal. This suggests that the current-induced change in anisotropy field alone cannot function as a recording mechanism. However, it is well-suited as a driving mechanism for parametric magnetization reversal\cite{ZayetsJMMM2023Parametric}. As a resonance-based method, parametric reversal is efficient and does not require a large change in the anisotropy field. The primary requirement for the driving mechanism of parametric reversal is that the polarity of the current-induced change should reverse when the current is reversed. Since the component shown in Figure \ref{fig:FigDistributionLinear} is linearly proportional to the current, it perfectly meets this requirement.

As shown in Fig. \ref{fig:FigDistributionLinear},  for the majority of nanomagnets,  a positive current-induced change in $k_{so}$ corresponds to a negative change in $H_{ani}$. Given that $H^0_{ani}$ and $k_{so}$  are not independent parameters, this observation is both unexpected and intriguing.  $H_{ani}$ and $k_{so}$ are related through the internal magnetic field $H_{int}$. In equilibrium, the magnetization is aligned along the magnetic easy axis by the internal magnetic field. Therefore, in the absence of an external field, the anisotropy field can be calculated from Eq. (\ref{EqHaniHz}) as:

\begin{equation}
	H_{ani}=k_{so} \cdot H_{int}
	\label{EqHaniHint}
\end{equation}

where in Eq. (\ref{EqHaniHz}) 

\begin{equation}
	H^0_{ani}= (1+k_{so})H_{int}
	\label{EqHaniHintkso}
\end{equation} 

As indicated in Eq. \ref{EqHaniHintkso},  $H_{ani}$  is linearly proportional to $k_{so}$, suggesting that their changes should have the same polarity. However, an exception occurs when another parameter, in addition to $k_{so}$,  influences $H_{ani}$ with a current dependency opposite to that of $k_{so}$

An additional intriguing feature of Fig. \ref{fig:FigDistributionLinear} is that all data points appear to align along a straight line, even for nanomagnets with different structures. This alignment can only be explained if this additional parameter influencing $H_{ani}$ is linearly proportional to $k_{so}$. This suggests that this additional parameter and $k_{so}$ are either directly or indirectly related to each other.

\begin{figure}[tb]
	\begin{center}
		\includegraphics[width=7.2 cm]{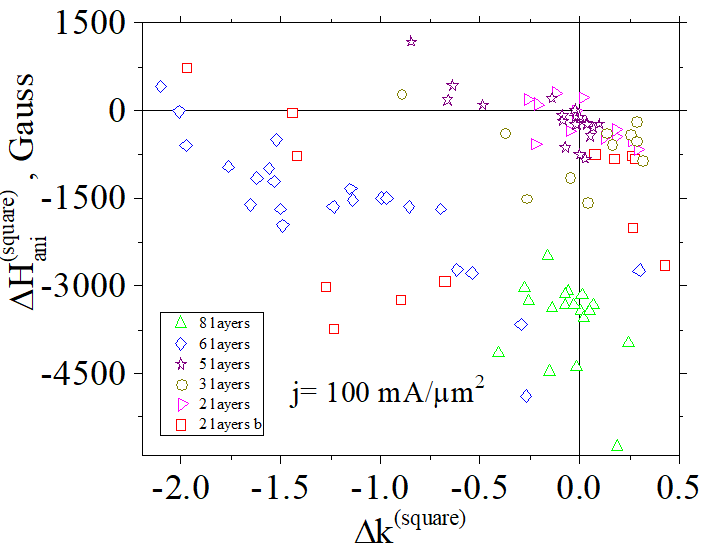}
	\end{center}
	\caption{
		{The change of anisotropy field  $\Delta H_{ani}^{(square)}$  versus the change of spin-orbit interaction coefficient $\Delta k_{so}^{(square)}$ as the current density increases from from 0 to 100 $mA/\mu m^2$  in multi-layer nanomagnets with varying numbers of layers. Only the components proportional to the square of the current are shown. Each dot represents an individual nanomagnet measurement. Dots of the same color and shape correspond to nanomagnets fabricated at different locations on the same wafer. } 
	}
	\label{fig:FigDistributionSq} 
\end{figure}

As shown in Figures \ref{fig:FigHaniKsoCurrentSingle} and \ref{fig:FigHaniKsoCurrentMulti}, the linear proportionality of $H_{ani}$ to the current is a feature exclusive to single-layer nanomagnets. In multi-layer nanomagnets, the primary component of current dependency is proportional to the square of the current (see Figure \ref{fig:FigHaniKsoCurrentMulti}). Figure \ref{fig:FigDistributionSq} illustrates the square-current component of the current- induced change in $H_{ani}$ versus the square-current component of the current- induced change in $k_{so}$ at the same current levels shown in Figure \ref{fig:FigDistributionLinear}.

The first noticeable feature is that the current-induced change in this case is an order of magnitude larger than that in a single-layer nanomagnet. Since in a multi-layer nanomagnet the current-induced change of the anisotropy field is proportional to the square of the current, it is independent of the current polarity. Therefore, the parametric mechanism of magnetization reversal cannot be used in multi-layer nanomagnets. However, the current-induced change of $H_{ani}$ is so large in this case, reaching 20-50$\%$, that it may itself trigger magnetization reversal through a thermo-activated mechanism \cite{ZayetsArch2019Hc}.

The probability of thermo-activated switching is exponentially proportional to the ratio of the magnetic energy of the nucleation domain to the thermal energy \cite{ZayetsArch2019Hc}. The magnetic energy equals the product of the magnetization and the internal magnetic field $H_{int}$. As follows from Eq. (\ref{EqHaniHintkso}), this means that the thermo-activated switching probability exponentially increases with a decrease in the anisotropy field. Therefore, the larger current- induced reduction in anisotropy shown in Figure \ref{fig:FigDistributionSq} may lead to thermo-activated switching and, consequently, to data recording.

In contrast to Figure \ref{fig:FigDistributionLinear}, there is no polarity anomaly between the current-induced changes in $H_{ani}$ and $k_{so}$ in Figure \ref{fig:FigDistributionSq}.  Nearly all nanomagnets show a negative $\Delta k_{so}^{(square)}$, which expectedly corresponds to a negative value of  $\Delta H_{ani}^{(square)}$. Additionally, unlike in Figure \ref{fig:FigDistributionLinear}, the data do not align along a single straight line. However, data from nanomagnets fabricated on the same wafer do align along individual lines. Similar to Figure \ref{fig:FigDistributionLinear}, the slope of these lines is negative and consistent across all wafers.

\section{Small Effect of Heating on  $j^2$-Contribution}

The component proportional to the square of the current can be associated with the heating of the nanomagnet. The heating energy is proportional to $j^2$,  and thus, the increase in temperature should also be proportional to $j^2$. This results in a corresponding decrease in the strength of the spin-orbit interaction and the anisotropy field. One might assume that the current-square contribution originates solely from heating, but this is not the case due to several observed features.

First, there is a difference in the polarity of $\Delta k_{so}^{(square)}$. In the case of the multilayer nanomagnet shown in Fig. \ref{fig:FigHaniKsoCurrentMulti}(b), $\Delta k_{so}^{(square)}$ is negative, indicating that spin-orbit strength decreases with heating. Conversely, in the single-layer nanomagnet shown in Fig. \ref{fig:FigHaniKsoCurrentSingle}(b), $\Delta k_{so}^{(square)}$ is positive, indicating that spin-orbit strength increases with heating. If $k_{so}$ were solely affected by heating, the polarity of its change should be consistent across all nanomagnets.

Second, there is a substantial difference in the magnitude of change in $H_{ani}$, At the same current density, and thus nearly the same heating temperature, the change in $H_{ani}$  for a multilayer nanomagnet is an order of magnitude larger than for a single-layer nanomagnet. Although the thermal conductivity properties may vary slightly between single-layer and multi-layer nanomagnets, these differences alone cannot account for the significant variation in the change of $k_{so}$ and $H_{ani}$ observed between the two types of nanomagnets. This suggests that the major contribution to the current-square components of $k_{so}$ and $H_{ani}$ originates from a mechanism other than heating.

The change in temperature was roughly estimated from the measured change in the resistance of the nanowire with the nanomagnet and the known thermo-resistance coefficients of the metals. The estimated heating does not exceed 100°C, which is still far below the Curie temperature of the studied nanomagnets, which is above 500°C. This further indicates that heating alone cannot account for the substantial changes observed in $k_{so}$ and $H_{ani}$.

The primary factor determining the temperature dependence of $k_{so}$ and $H_{ani}$  is the nanomagnet's magnetization. It has been observed that modulating the magnetization via gate voltage results in a modulation of $H_{ani}$ \cite{ZayetsArch2024_SO_VCMA}. The strength of the spin-orbit interaction is proportional to the intrinsic magnetic field in a nanomagnet, which in turn is proportional to the magnetization. A rise in temperature reduces magnetization and, consequently, $k_{so}$.  According to Eq. \ref{EqHaniHint}, the reduction of $k_{so}$ causes a reduction in $H_{ani}$. As Figure \ref{fig:FigDistributionSq} shows, $\Delta H_{ani}^{(square)}$ is mostly negative, but $\Delta k_{so}^{(square)}$ can be both negative and positive. These inconsistent polarities can only be explained by the existence of a temperature-independent mechanism with a more complex contribution to the current-square changes in $k_{so}$ and $H_{ani}$.

\section{ Modulation of Magnetic Anisotropy and Spin-Orbit Interaction Induced by Spin Accumulation}

The mechanisms by which electrical current modulates spin-orbit interaction and anisotropy are not entirely clear, but it is evident that spin accumulations created by the current, particularly their associated magnetic fields, play a crucial role. A magnetic field is generated along the spin direction of spin-polarized electrons. The component of this field that is perpendicular to the magnetic easy axis causes a tilt in the magnetization, which can be measured through this deflection \cite{ZayetsJMMM2023Parametric}. Although this component is typically small, under certain conditions, it can be used to reverse magnetization via the parametric resonance mechanism  \cite{ZayetsJMMM2023Parametric}. 

It can be suggested  that there is also a component of the magnetic field from spin-accumulated electrons that is aligned along the magnetic easy axis. Unlike the perpendicular component, this field does not tilt the magnetization but instead affects the strength of the magnetic anisotropy and spin-orbit interaction. Measuring or even distinguishing this parallel component is challenging because, even in equilibrium, a substantial internal magnetic field, $H_{int}$, exists along the easy axis. This internal field, which is responsible for the existence of the magnetic anisotropy and strong spin-orbit interaction in equilibrium \cite{Zayets2024SObasic}, is much larger and can obscure the smaller magnetic field generated by spin-accumulated electrons.

Since both $k_{so}$ and $H_{ani}$ are proportional to $H_{int}$, any additional magnetic field component aligned with $H_{int}$ would proportionally increase their magnitudes. Therefore, it is crucial to note that only the component of the magnetic field from spin-accumulated electrons that is aligned with $H_{int}$ enhances magnetic anisotropy and spin-orbit interaction, while the perpendicular component merely tilts the magnetization without impacting anisotropy.

\section{Identifying and Differentiating Between Two Sources of Spin Accumulation}

The first effect leading to spin accumulation at the nanomagnet interfaces is the Spin Hall effect \cite{SpinHall_Kato2004,SpinHall_Dyakonov1971,SpinHall_Sinolva2005}. Initially, the spins accumulated due to the Spin Hall effect are oriented in-plane and do not impact the anisotropy or spin-orbit interaction. However, due to spin precession  along the internal magnetic field and associated precession damping, these spins gradually align along the easy magnetic axis and, consequently, begin to influence both anisotropy and spin-orbit interaction. This spin precession and alignment have been experimentally observed \cite{ZayetsJMMM2023Parametric}, showing that the alignment is relatively rapid. This demonstrates that a significant portion of spin accumulation from the Spin Hall effect does, in fact, contribute to the modulation of anisotropy and spin-orbit interaction.

The second effect contributing to spin accumulation is the Ordinary and Anomalous Hall effects. The Hall effect is relatively weak in metals compared to semiconductors; for example, Hall voltage in metals is about three orders of magnitude smaller than in semiconductors \cite{HallMetal}. This is because metals have roughly equal numbers of electrons and holes, whose opposite contributions to the Hall effect nearly cancel each other out \cite{ZayetsJMMM2018Holes}. However, a slight imbalance between the numbers of electrons and holes creates some Hall voltage. The accumulation of large, but nearly equal numbers of negatively charged electrons and positively charged holes on the same side of the nanomagnet due to the Hall effect results in minimal net charge accumulation, leading to a small Hall voltage. However, this phenomenon generates significant spin accumulation in ferromagnetic metals for the following reason.

In ferromagnetic metals, conduction electrons are spin-polarized with spin directed along the magnetization. Despite their opposite charges, spin-polarized electrons and holes share the same spin direction, aligning with the magnetization. Consequently, when spin-polarized electrons and holes accumulate on one side of the nanomagnet, they are collectively adding their spins to enlarge the spin accumulation.

It is essential to emphasize two key differences between the spin accumulation mechanisms induced by the Spin Hall effect and those resulting from the Ordinary (or Anomalous) Hall effects. Firstly, the Spin Hall effect can generate spin accumulation in both ferromagnetic and non-magnetic metals because it converts spin-unpolarized electrons into spin-polarized electrons, making the pre-existence of spin-polarized electrons unnecessary. In contrast, the Ordinary (or Anomalous)  Hall effect does not create new spins; instead, they redistribute existing spin-polarized electrons, which means that their initial presence is required.

Secondly, the Spin Hall effect produces spins with an in-plane orientation perpendicular to the current flow, dictated by the current direction and the symmetry of spin-orbit interaction.  In contrast, the spin accumulation resulting from the Ordinary and Anomalous Hall effects aligns along the pre-existing spins, which means they are oriented along the magnetization and, in the studied nanomagnets, perpendicular to the plane.

\section{Experimental Identification and Separation of Two Contributions}

Each contribution can be experimentally distinguished based on the current-dependence of the modulation of $H_{ani}$ and $k_{so}$. The contribution from the Ordinary (or Anomalous) Hall effect is linearly proportional to the current, while the contribution from the Spin Hall effect is quadratically proportional to the current. All effects—Spin Hall, Ordinary Hall, and Anomalous Hall—are linearly proportional to the current and change polarity when the current is reversed. Based on this alone, one might incorrectly conclude that both mechanisms contribute only to the component linearly proportional to the current. However, the post-creation alignment of the spin accumulation along the magnetization adds complexity to these dependencies.

The spin accumulation created by the Ordinary and Anomalous Hall effects is always aligned along the magnetization, and therefore, there is no post-creation spin realignment. As a result, the contributions of this mechanism to the modulation of  $H_{ani}$ and $k_{so}$ remains linearly proportional to the current.

In contrast, the spin accumulation resulting from the Spin Hall effect contributes only to the component proportional to the current squared. This can be inferred from how the polarity of the contribution behaves when the current is reversed. Since this contribution does not change polarity with current reversal, it behaves as an even function of the current, making a current-squared dependence the most likely scenario due to its lowest order nature.

The reason why the Spin Hall effect's contribution to the modulation of anisotropy and spin-orbit interaction is independent of current polarity, even though the generated spin direction reverses with current reversal, can be explained as follows. The spins generated by the Spin Hall effect are initially oriented in-plane, perpendicular to both the magnetization and the spin direction of the existing spin-polarized conduction electrons.  Immediately after the spins are accumulated, the internal magnetic field begins to align these spins along the magnetization. Although the initial in-plane spin direction is reversed when the current is reversed, it remains in-plane, and the final after- alignment spin direction is always aligned with the magnetization regardless of the initial in-plane spin direction. This means that the component of the spin accumulation created by the Spin Hall effect, which affects $H_{ani}$ and $k_{so}$,  is always ultimately aligned along the magnetization, independent of the current polarity and its initial spin direction.  This persistent alignment along the magnetization, independent of current polarity, results in the spin accumulation due to the Spin Hall effect contributing solely to the component proportional to the current squared.

\section{Enhanced Effect in Multilayer Nanomagnets}

Spin accumulation due to the Spin Hall, Anomalous Hall, and Ordinary Hall effects occurs only at the interfaces \cite{SpinHall_Kato2004} between two different materials. Since the current-induced modulation of magnetic anisotropy and spin-orbit interaction is proportional to the total amount of spin accumulation, increasing the number of interfaces, as in a multilayer nanomagnet, could potentially amplify the effect. For instance, adding a thin ferromagnetic layer to a nanomagnet introduces two new interfaces—one above and one below the layer—creating additional sites for spin accumulation, which could enhance the modulation effect. However, this enhancement occurs only when the contributions from both interfaces share the same polarity, thus adding to each other, rather than canceling each other out if their polarities are opposite. By comparing contributions from opposite sides of the inserted layer, it is possible to determine whether increasing the number of layers in a nanomagnet will indeed lead to an enhancement of the effect or not.

The polarity of spin accumulation reverses on opposite sides of a nanomagnet, regardless of the type of Hall effect responsible for it. The Ordinary (or Anomalous) Hall mechanism merely redistributes the spins, causing an increase in spins at one location and a corresponding decrease at another.  As a result, spin accumulation at one interface directly corresponds to spin depletion at the opposite interface. This leads to a cancellation effect: any current-induced change in the strength of the spin-orbit interaction at one interface is fully offset by the opposite change at the other.  Therefore, the insertion of an additional thin ferromagnetic layer does not enhance the component of the modulation of $H_{ani}$ and $k_{so}$ that is linearly proportional to the current. This feature was experimentally observed in the modulation of $H_{ani}$ and $k_{so}$, which are linearly proportional to the current (See Fig. \ref{fig:FigHaniKsoCurrentSingle},\ref{fig:FigHaniKsoCurrentMulti}).

The influence of additional interfaces on spin accumulation due to the Spin Hall effect and its modulation of $H_{ani}$ and $k_{so}$  is substantially different. Even though the spin direction is opposite on the two sides of an inserted layer for spin accumulation created by the Spin Hall effect \cite{SpinHall_Kato2004}, this does not lead to any cancellation or reduction of the contribution from each interface. This is because the Spin Hall effect generates spins perpendicular to the magnetization at both interfaces, and these spins are subsequently aligned along the same magnetization direction. Despite the initial opposite spins at opposite sides of the nanomagnet, they align along the same direction, resulting in the same polarity and an additive effect from each interface. Consequently, the effect accumulates with an increasing number of interfaces, as experimentally observed in the modulation of $H_{ani}$ and $k_{so}$, which are proportional to the current squared (see Figs. \ref{fig:FigHaniKsoCurrentMulti},\ref{fig:FigDistributionSq}).

\section{Metal-Metal vs. Metal-Dielectric Interfaces}

The amount of spin accumulation, and consequently the magnitude of current-induced modulation of $H_{ani}$ and $k_{so}$, is strongly influenced by the type of interface—whether it is a metal-metal or metal-dielectric interface—and by the specific materials on each side of the interface.

The first key factor is spin damping near the interface: even with the same spin pumping due to Hall effects, variations in spin damping result in different amounts of spin accumulation.

The second factor is the spatial distribution of accumulated spins. A broader spread of spin accumulation increases spin damping. Additionally, in thin layers, a broad spatial distribution may lead to interactions between spin accumulation on one side and spin depletion at the opposite interface, effectively reducing the net spin accumulation and depletion on each side.

 The behavior of spin accumulation varies depending on whether the material at the interface is an insulator or a metal. When one part of the interface is an insulator, the region of spin accumulation is small and localized precisely at the interface, thus contributing significantly to the current-induced modulation of $H_{ani}$ and $k_{so}$.  In contrast, at a metal-metal interface, charge and spin can flow through the interface, causing spins to diffuse from the interface deep into the non-magnetic metal. This results in substantial spin relaxation and a significant reduction in spin accumulation. Therefore, the contribution of a metal-metal interface to current-induced modulation of $H_{ani}$ and $k_{so}$ is smaller.

\section{Symmetrical vs. Asymmetrical  Nanomagnets}

As described above, the current-induced modulation of magnetic anisotropy and spin-orbit interaction is attributed to the magnetic field generated by current-induced spin accumulation. This perpendicular-to-plane magnetic field is directly proportional to the total spin accumulation in a nanomagnet, regardless of its spatial distribution across the nanomagnet's thickness.

The Ordinary (or Anomalous) Hall effect does not alter the total spin in a nanomagnet, as it merely spatially redistributes existing spins rather than creating new ones. Consequently, in symmetrical nanomagnets (e.g., MgO/Fe/MgO), the contribution that is linearly proportional to the current is absent. However, in an asymmetrical nanomagnet (e.g., Ta/Fe/MgO), the different spin relaxation rate at the two opposing interfaces imbalances the spin accumulation and spin depletion,  effectively altering the total spin of the nanomagnet and resulting in existence of a contribution that is linearly proportional to the current.

Similarly, the Spin Hall effect does not affect the total spin of the nanomagnet because it generates equal amounts of spins in opposite directions at opposite interfaces. However, these spins are subsequently realigned in the same direction due to the internal magnetic field. As a result, there is no cancellation of contributions from opposite sides of the nanomagnet and a contribution of the $H_{ani}$ and $k_{so}$ modulation, which is  proportional to the current squared, exists in both symmetrical and asymmetrical nanomagnets.

\section{Conclusion}

In conclusion, we have observed and measured the modulation of magnetic anisotropy and the strength of spin-orbit interaction in a nanomagnet induced by an electrical current passing through it. Our systematic experimental study revealed two distinct contributions to this modulation: one proportional to the current and the other proportional to the square of the current. The squared-current contribution originates from the Spin Hall effect, which uniquely accumulates in strength with an increasing number of interfaces. Consequently, the current modulation of magnetic anisotropy and spin-orbit interaction is exceptionally large in multi-layer nanomagnets. For current densities typically used for MRAM recording, the change in the anisotropy field can reach 30-50$\%$, leading to thermally-activated magnetization reversal and data recording.

The Ordinary and Anomalous Hall effects are responsible for the contribution linearly proportional to the current. This contribution has opposite polarity at opposite interfaces of the nanomagnet and thus exists only in asymmetrical nanomagnets with different interfaces. It does not accumulate with an increased number of interfaces, making it small and significant only in single-layer nanomagnets. This small modulation of the magnetic anisotropy is insufficient to cause thermally-activated magnetization reversal. However, since the polarity of the current modulation of anisotropy follows the polarity of the current, it can be utilized in parametric magnetization reversal \cite{ZayetsJMMM2023Parametric}. As a resonance-based reversal method, it does not require large modulation of magnetic anisotropy, and the measured modulation is sufficient for this type of magnetization reversal.

These findings offer critical insights into the diverse behaviors of current-induced spin accumulation and its impact on spin-orbit interaction and magnetic anisotropy at the nanoscale. This knowledge paves the way for optimizing recording methods, reducing recording currents, and developing innovative data recording schemes for magnetic storage and spintronic devices. Since all types of magnetic memory rely on magnetic anisotropy, comprehending how it is influenced by electrical current can significantly enhance and advance various magnetic memory technologies.

\bibliographystyle{elsarticle-num} 

\bibliography{SOT_SpinOrbitBib}

\end{document}